# On an Extension of Stochastic Approximation EM Algorithm for Incomplete Data Problems

## Vahid Tadayon[1]

**Abstract**: The Stochastic Approximation EM (SAEM) algorithm, a variant stochastic approximation of EM, is a versatile tool for inference in incomplete data models. In this paper, we review fundamental EM algorithm and then focus especially on stochastic version of EM. In order to construct the SAEM, the algorithm combines EM with a variant of stochastic approximation that uses Markov chain Monte-Carlo to deal with the missing data. The algorithm is introduced in general form and can be used to a widely range of problems.

**Keywords**: Stochastic Approximation; EM algorithm; Incomplete Data; Markov chain Monte-Carlo; Maximum Likelihood.

1. ## Introduction

A standard method to handle incomplete data problems is EM algorithm (Dempster et al., 1977; Tadayon and Torabi, 2018). This procedure is an iterative method to find maximum likelihood in some incomplete data. This algorithm is applied to widely various problems. However, the convergence of this method can be slow. Furthermore, in situations where the data are dependent and incomplete (for example, spatial incomplete data problems) this method can be highly inefficient. To resolve some of these difficulties, we explore Stochastic Approximation EM (SAEM). To pay this algorithm, first Stochastic Approximation (SA) is considered. Stochastic Approximation (SA) was introduced by Robbins and Monro (1951), and has been extended by Gu and Kong (1998) to situations where data are incomplete. We utilize SA with MCMC to construct SAEM. SAEM algorithm originates from Delyon et al. (1999). In this paper, we propose an extension form of SAEM based on SA with MCMC (Gu and Kong, 1998).

In the next section, EM algorithm is reviewed. Then SAEM is introduced.

2. ## EM algorithm

We assume that x is observed (or incomplete) data and is generated by some distribution. Let z denote the unobserved (or missing) data. Hence in EM algorithm pair (x,z) is recognized as complete data. Let $f(x,z;\theta)$ denote the joint distribution of the complete data, dependent on parameter vector $\theta$. With this new density function, we can define a new likelihood function $L(\theta) = L(\theta;x) = \int f(x,z;\theta)dz$, which is referred to as the incomplete-data likelihood function. The goal is to find $\hat{\theta}$, the maximize of the marginal likelihood $L(\theta)$.

---
[1]Department of Statistics, Tarbiat Modares University, Tehran-Iran.

The EM algorithm consisted of two stage. First computes the expected value of the complete-data log-likelihood $\log f(x,z;\theta)$ with respect to the unknown data z given the observed data x and the current parameter estimates.

That is, we define:

$$Q(\theta, \theta^{(i-1)}) = E[\log f(x,z;\theta) | x, \theta^{(i-1)}] \qquad (1)$$

where $\theta^{(i-1)}$ are the current parameters estimates. Second, we maximize the expectation.

This is the M-step. Given an initial value $\theta^{(0)}$, the EM algorithm produces a sequence $\{\theta^{(0)}, \theta^{(1)}, \theta^{(2)}, ...\}$ that, under mild regularity conditions (Boyles, 1983), converges to $\hat{\theta}$.

To end this section, we mention some challenges for the EM algorithm. One of the biggest challenges for the EM algorithm is that it only guarantees convergence to a local solution (Jank, 2006; Tadayon and Rasekh, 2018; Tadayon, 2017; Tadayon and Khaledi, 2015; Tadayon, 2018). The EM algorithm is a grasping method in the sense that it is attracted to the solution closest to its starting value. Then the next problem with EM algorithm is starting values. In addition, in some cases the likelihood function is computationally intractable and it is infeasible to maximize the likelihood function of observed data directly.

To avoid above problems, in the next section SAEM is introduced.

## 3. Stochastic Approximation EM algorithm

Using likelihood function $L(\theta)$, the maximum likelihood estimate of $\theta$, denoted by $\hat{\theta}$, is defined by

$$L(\hat{\theta};x) = \max_{\theta} L(\theta, x). \qquad (2)$$

Due to computationally intractable (2), we consider the first-order and second- order partial derivatives of the log-likelihood function in order to use gradient-type algorithms, such as Newton-Raphson and Gauss-Newton algorithms (Ortega, 1990).

### 3.1 Derivatives of the log-likelihood function

The first –order and second-order derivatives of the log-likelihood functions can be derived by using the log-likelihood functions of complete data, denoted by $l_c(\theta;x,z)$.

From the missing information principle, the first-order derivative of $L(\theta;x)$, called the score function, can be written as

$$s_\theta(\theta;x) = \partial_\theta \log L(\theta;x) = E[S_\theta(\theta;z) | x, \theta], \qquad (3)$$

where $S_\theta(\theta;z) = \partial_\theta l_c(\theta;x,z)$ and $E[. | x, \theta]$ denotes that the expectation is taken with respect to the conditional distribution $f(z | X = x, \theta)$. In addition, we use $\partial$ and $\partial^2$ to denote the first-order and second-order derivatives with respect to a parameter vector, say

$\partial_\theta a(\theta) = \partial a(\theta)/\partial \theta$ and $\partial^2_\theta a(\theta) = \partial^2 a(\theta)/\partial \theta \partial \theta^T$. To calculate the second order derivative of the log-likelihood function, we apply Louis's (1982) formula to obtain

$$-\partial^2_\theta \log L(\theta; x) = E[I_{\theta\theta}(\theta; z) - S_\theta(\theta; z)^{\otimes 2} | x, \theta] + s_\theta(\theta, x)^{\otimes 2}, \quad (4)$$

where for vector $a, a^{\otimes 2} = aa^T$ and $I_{\theta\theta}(\theta; z) = -\partial^2_\theta l_c(\theta; x, z)$ denotes the information matrix for complete data.

## 3.2 Steps of the SAEM algorithm

At the i-th iteration, $\theta^{(i)}$ is the current estimate of $\hat{\theta}$; $h^{(i)}$ the current estimate of $s_\theta(\hat{\theta}; x)$; $\Gamma^{(i)}(t)$, the current estimate of $E[I_{\theta\theta}(\hat{\theta}; z) - tS_\theta(\hat{\theta}; z)^{\otimes 2} | x, \hat{\theta}] + s_\theta(\hat{\theta}, x)^{\otimes 2}$. We assume that $\Pi_{x,\theta}(.,.)$ is the transition probability of the Metropolis-Hastings algorithm used to simulate from the conditional distribution of z given x and $\theta$.

Step1. At the i-th iteration, set $z^{(i,0)} = z^{(i-1,N_{i-1})}$. Generate $z^{(i)} = (z^{(i,1)}, ..., z^{(i,N_i)})$ from the transition probability $\Pi_{x,\theta^{(i-1)}}(z^{(x,i,k-1)},.)$.

Step2. Update the estimates as follows:

$$\theta^{(i)} = \theta^{(i-1)} + \gamma_i [\Gamma^{(i)}(t)]^{(-1)} \bar{H}(\theta^{(i-1)}; z^{(x,i)})$$
$$h^{(i)} = h^{(i-1)} + \gamma_i (\bar{H}(\theta^{(i-1)}; z^{(x,i)}) - h^{(i-1)})$$
$$\Gamma^{(i)} = \Gamma^{(i-1)} + \gamma_i (\bar{I}(\theta^{(i-1)}; z^{(x,i)}) - \Gamma^{(i-1)})$$

where $t \in [0,1]$,

$$\bar{I}(\theta, z^{(x,i)}) = \frac{1}{N_i} \sum_{k=1}^{N_i} I_{\theta\theta}(\theta, z^{(x,i,k)})$$

$$\bar{H}(\theta, z^{(x,i)}) = \frac{1}{N_i} \sum_{k=1}^{N_i} S_\theta(\theta, z^{(x,i,k)})^T.$$

Finally, the constants sequence $\{\gamma_i\}$ satisfies the following conditions:

$$0 \leq \gamma_i \leq 1 \text{ for all i, } \sum_{i=1}^\infty \gamma_i = \infty \text{ and } \sum_{i=1}^\infty \gamma_i^2 < \infty.$$

An important feature of the SAEM algorithm is that it uses a constants sequence $\{\gamma_i\}$ to handle the noise in approximating $\partial_\theta \log L(\theta; x)$ and $\partial^2_\theta \log L(\theta; x)$ in Step2 (Robbins and Monro, 1951; Lai, 2003).

Now, we consider the convergence of the algorithm. In order to achieve this goal, it can be shown that the sequence of parameters estimates of $\theta$ returned by SAEM algorithm, approximate the solution to the differential equation

$$\frac{d}{di}\theta(i) = \Gamma^{-1}(i)E[\bar{H}(\theta,z)|x,\theta]$$

with the corresponding terms for the $\Gamma^{(i)}(t)$. For more details and conditions of this convergence see theorem 3.12 of Benveniste et al. (1990) and Gu and Kong (1998). It is sufficient to check out that the distribution under study satisfies the conditions of theorem 3.12 of Benveniste et al. (1990).

4. Conclusion

In this study, we raise a stochastic approximation interpretation for EM algorithm. This stochastic approximation viewpoint provides some convenience for EM algorithm. It also suggests a more flexible way to maximization step of EM algorithm by using MCMC. It should be emphasized that the main goal of the current paper is concentration on the role of stochastic approximation in expectation stage of EM algorithm.